\documentclass[12pt]{article}
\usepackage{a41}
\usepackage{rotating}
\usepackage{epsfig}
\usepackage{psfig}
\usepackage{amsmath,bbm}
\usepackage{color}
\begin{document}

\newcommand{\be}{\begin{equation}}
\newcommand{\ee}{\end{equation}}

\newcommand{\cc}{\cite}
\newcommand{\ba}{\begin{eqnarray}}
\newcommand{\ea}{\end{eqnarray}}

\begin{titlepage}
\begin{center}

\vspace{5cm}

  {\Large \bf Anomalous Quark Chromomagnetic Moment and Single-Spin Asymmetries }
  \vspace{0.50cm}\\
 Nikolai Kochelev \footnote{kochelev@theor.jinr.ru} and  Nikolai Korchagin \footnote{korchagin@theor.jinr.ru}
 \\
{ \it Bogoliubov Laboratory of Theoretical Physics, Joint
Institute for Nuclear Research, Dubna, Moscow region,
141980, Russia}\\
 \vskip 1ex
\end{center}
\vskip 0.5cm \centerline{\bf Abstract}

We discuss a nonperturbative  mechanism for the single-spin
asymmetries  in the strong interaction. This mechanism is based on
the existence of a large anomalous quark chromomagnetic moment
induced by the  nontrivial topological  structure of QCD vacuum.
Our estimations within the instanton liquid model for QCD vacuum
show that AQCM generates very large SSA on the quark level.
Therefore, this mechanism can be responsible for the anomalously
large SSA observed in different high energy reactions with
hadrons.

\end{titlepage}

\setcounter{footnote}{0}

\section{Introduction}

One of the longstanding problems in strong interaction is the
understanding within  QCD of the mechanism that is responsible for
the large  single-spin asymmetries (SSA) observed  in numerous
high energy reactions with hadrons. Many different approaches were
suggested to solve this problem (see recent papers and review
\cite{Anselmino:2013rya,Anselmino:2012rq,reviews} and references
therein). Most of them are based on the assumption of the
so-called transverse-momentum-dependent (TMD) factorization
\cite{Ji:2004xq,Ji:2004wu,Bacchetta:2008xw,collins}. The validity
of this assumption  is not clear so far \cite{Rogers:2010dm}.
Furthermore, in our paper we will show the existence of the
nonperturbative QCD mechanism which violates explicitly the TMD
factorization for SSA.

It is well known that SSA arises from interference of different
diagrams and should include at least two ingredients. First off
all, it should be a helicity-flip in the scattering amplitude and
secondly, the amplitude should have a nonzero imaginary part. The
small current masses of quarks are only a source in perturbative
QCD (pQCD) for helicity-flip. Furthermore, the imaginary part of
the scattering amplitude, which comes from loop diagrams, is
expected to be suppressed by extra power of the strong coupling
constant $\alpha_s$. As a result, pQCD fails to describe large
observed SSA. On the other hand, it is known that QCD has a
complicated structure of vacuum which leads to the phenomenon  of
spontaneous chiral symmetry breaking (SCSB) in strong interaction.
Therefore, even in the case of a very small current mass of the
quarks their dynamical masses arising  from SCSB  can be  large.
The instanton liquid model of QCD vacuum \cite{shuryak,diakonov}
is one of the models in which the SCSB phenomenon arises  in a
very natural way due to quark chirality flip in the field of
strong
 fluctuation of the vacuum gluon field called
instanton \cite{Belavin:1975fg,'tHooft:1976fv}.
 The instanton is the well-known solution of QCD equation of motion in the Euclidian space-time which
has   nonzero topological charge. In many papers (see reviews
\cite{shuryak,diakonov,Kochelev:2005xn}), it was shown that
instantons play a very important role in hadron physics.
Furthermore, instantons lead to the anomalous quark-gluon
chromomagnetic  vertex with a large
  quark helicity-flip  \cite{kochelev1,diakonov}.
Therefore, they can give the important contribution to SSA
\cite{kochelev1,Kochelev:1999nd,Cherednikov:2006zn,Dorokhov:2009mg,Ostrovsky:2004pd,diakonov,Qian:2011ya}.

In this paper, we will present  the  first consistent calculation
of SSA in the quark-quark scattering based on the existence of the
anomalous quark chromomagnetic moment (AQCM) induced by instantons
\cite{kochelev1} \footnote{ The semi-classical mechanism for SSA
based on large AQCM has recently been discussed  in  papers
\cite{Abramov:2011zz,Abramov:2009tm}.}.

\section{ Quark-gluon interaction in non-perturbative QCD }

In the  general case, the interaction vertex of a massive quark
with a gluon, Fig.1, can be written in the following form:
\begin{equation}
V_\mu(p_1^2,{p_1^\prime}^2,q^2)t^a = -g_st^a[F_1(p_1^2,{p_1^\prime}^2,q^2) \gamma_\mu
 +
\frac{\sigma_{\mu\nu}q_\nu}{2M_q}F_2(p_1^2,{p_1^\prime}^2,q^2)],
 \label{vertex}
 \end{equation}
where the first term is  the conventional perturbative QCD
quark-gluon vertex and the second term comes from the
nonperturbative sector of QCD.
\begin{figure}[htb]
\vspace*{2.0cm} \centering
\centerline{\epsfig{file=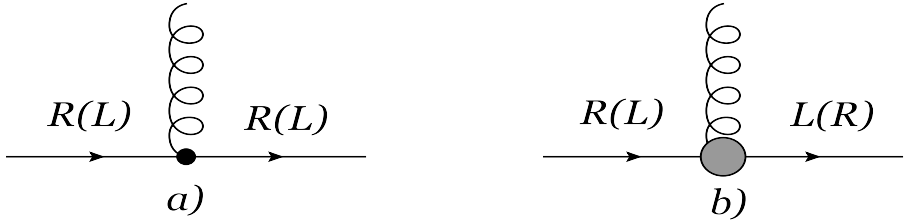,width=12cm,height=3.0cm,
angle=0}} \vskip 1cm \caption{a) Perturbative helicity non-flip
and b)  nonperturbative helicity-flip quark-gluon vertices}
\label{vertexpicture}
\end{figure}

In Eq.\ref{vertex} the form factors $F_{1,2}$ describe
nonlocality of the interaction, $p_{1}, p_1^\prime$ are
the momenta of incoming and outgoing quarks, respectively, $ q=p_1^\prime-p_1$,
 $M_q$ is the quark mass, and $\sigma_{\mu\nu}=(\gamma_\mu \gamma_\nu-\gamma_\nu \gamma_\mu)/2$.

The form factor  $F_2(p_1^2,{p_1^\prime}^2,q^2)$
 suppresses the AQCM vertex
at short distances when the respective virtualities are large. Within the
instanton model it is  explicitly  related to the  Fourier-transformed quark
zero-mode and instanton fields and reads
\begin{equation}
 F_2(p_1^2,{p_1^\prime}^2,q^2) =\mu_a
\Phi_q(\mid p_1\mid\rho/2)\Phi_q(\mid p_1^\prime\mid\rho/2)F_g(\mid
q\mid\rho) \ , \nonumber
\end{equation}
where
\begin{eqnarray}
\Phi_q(z)&=&-z\frac{d}{dz}(I_0(z)K_0(z)-I_1(z)K_1(z)), \label{ffq}\\
F_g(z)&=&\frac{4}{z^2}-2K_2(z), \label{ffg}
\end{eqnarray}
$I_{\nu}(z)$, $K_{\nu}(z)$ are the modified Bessel functions and
$\rho$  is the instanton size.

We assume  $F_1\approx 1$ and $F_2(p_1^2,{p_1^\prime}^2,q^2)
 \approx \mu_a F_g(q^2)$ since valence quarks in hadrons have small virtuality.

Within the instanton liquid model \cite{shuryak,diakonov}, where
all instantons have the same size $\rho_c$, AQCM is
\cite{kochelev2}
\begin{equation}
\mu_a=-\frac{3\pi (M_q\rho_c)^2}{4\alpha_s}.
\label{AQCM}
\end{equation}
 In  Eq.(\ref{AQCM}), $M_q$ is the so-called dynamical quark mass.
We would like to point out two specific features of the formula
for AQCM. First, the strong coupling constant enters into the
denominator  showing a clear nonperturbative origin of AQCM. The
second feature is the negative sign of AQCM. As we will see below,
the sign of AQCM leads to the definite sign of SSA in the
quark-quark scattering. The value of AQCM strongly depends on the
dynamical quark mass which   is $M_q=170$ MeV   in the mean field
approximation (MFA)\cite{shuryak}
 and $M_q=350$ MeV in the Diakonov-Petrov model (DP)   \cite{diakonov}.
Therefore,  for fixed value of the strong coupling constant  in
the  instanton model,  $\alpha_s\approx \pi/3\approx 0.5$
\cite{diakonov}, we get

\begin{equation}
{\mu_a}^{MFA}=-0.4 \  \ \   \mu_a^{DP}= -1.6
\label{mu}
\end{equation}

We would like to mention that the Schwinger-type   of the  pQCD
contribution to AQCM
\begin{equation}
    \mu_a^{pQCD}=-\frac{\alpha_s}{12\pi}\approx 1.3\cdot 10^{-2}
\label{pQCD}
\end{equation}
is by several orders of magnitude smaller in comparison with the
nonperturbative contribution induced by instantons, Eq.\ref{mu},
and, therefore, it  can give only a tiny contribution to
spin-dependent cross sections
  \cite{Ryskin:1987ya}\footnote{Recently, a rather large AQCM has been obtained within the approach based on the
 Dyson-Schwinger equations (see review \cite{Roberts:2012sv} and references therein).}.

\section{Single-spin asymmetry in high energy quark-quark scattering induced by AQCM}

The SSA for the process  of transversely polarized  quark  scattering  off unpolarized quark,
 $q^{\uparrow}(p_1)+q(p_2) \to q (p_1^\prime)
+ q(p_2^\prime)$, is defined as
\begin{equation}
A_{N}=\frac{d \sigma^{\uparrow} - d \sigma^{\downarrow}}
{d \sigma^{\uparrow}+d \sigma^{\downarrow}},
\end{equation}
where   ${\uparrow}  {\downarrow}$ denote the initial quark spin
orientation perpendicular to the scattering plane and

\begin{equation}
d \sigma^{\uparrow\downarrow}=\frac{|M(\uparrow\downarrow)|^2}{\mathrm{2I}} d\mathrm{PS_2(S,q_t)},
\end{equation}
where $I$ is the initial flux, $S=(p_1+p_2)^2$,
$M(\uparrow\downarrow)$ is the matrix element for the different
initial spin directions,  $d\mathrm{PS_2(S,q_t)}$ is the
two-particle phase space and $q_t={{p_1}^\prime}_t-{p_1}_t$ is the
transverse momentum transfer. In the high energy limit $S\gg
q_t^2,M_q^2$, we have $I\approx S$ and
$d\mathrm{PS_2(S,q_t)}\approx d^2q_t/(8\pi^2S)$.

In terms of the  helicity amplitudes \cite{Goldberger:1960md},
\cite{Buttimore:1978ry}
\begin{equation}
\Phi_1=M_{++;++},\ \   \Phi_2=M_{++;--},\ \   \Phi_3=M_{+-;+-},\ \   \Phi_4=M_{+-;-+}  ,\ \   \Phi_5=M_{++;+-},
\nonumber
\end{equation}
where the symbols $+$ or $-$ denote the helicity of quark in the
c.m. frame, SSA is given by
\begin{equation}
A_N=-\frac{2Im[( \Phi_1+\Phi_2+\Phi_3-\Phi_4)\Phi_5^*]} {|\Phi_1|^2+|\Phi_2|^2+|\Phi_3|^2+|\Phi_4|^2+4|\Phi_5|^2)}.
\label{helicity}
\end{equation}
In  Fig.2, we present the set of   diagrams which give a
significant contribution to $A_N$. Higher order terms  in $\mu_a$
and $\alpha_s$  are expected to be suppressed by a small instanton
density in QCD vacuum \cite{shuryak}  and by a large power of the
small strong coupling constant.

\begin{figure}[htb]

   \vspace*{-0.0cm} \centering
   \centerline{\epsfig{file=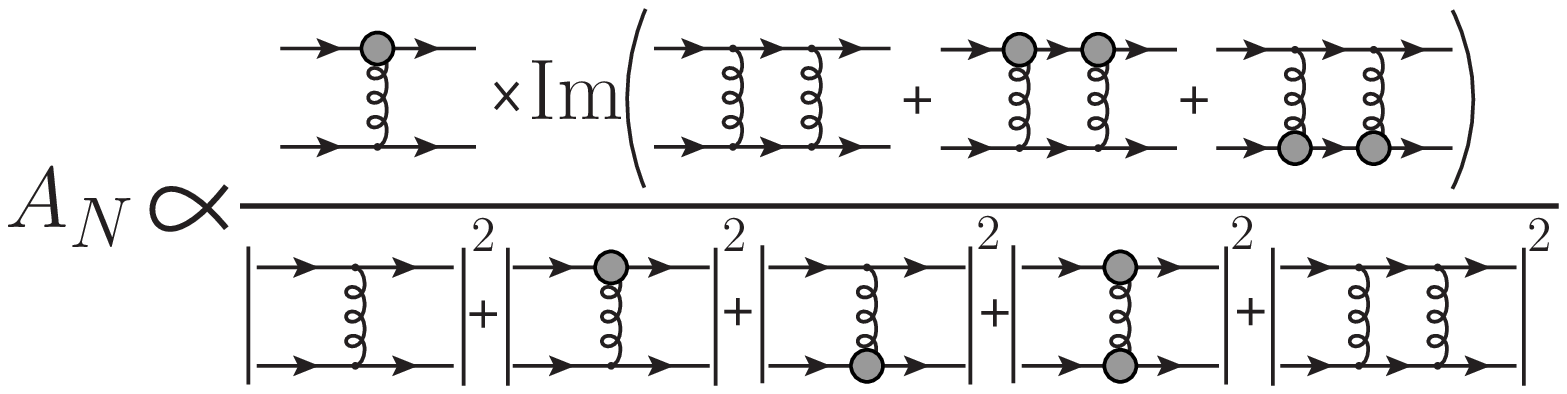,width=14cm, angle=0}}
    \caption{Contribution to SSA arising from different diagrams.}

   \end{figure}

For estimation, we take   the simple form for the  gluon propagator in the  Feynman gauge
         \begin{equation}
             P_{\mu\nu}(k^2)=\frac{g_{\mu\nu}}{k^2-m_g^2},
     \nonumber
       \end{equation} where $m_g$ can be treated as the   infrared cut-off related to confinement  \cite{nikolaev},
or as the dynamical gluon mass \cite{RuizArriola:2004en},
\cite{aguilar}.   Within the instanton model this parameter can be
considerated as the effect of
 multiinstanton contribution to the gluon propagator.

By using in the high energy limit the Gribov decomposition for the
metric tensor into the transverse and longitudinal parts
\begin{equation}
     g_{\mu\nu}=g_{\mu\nu}^t+ \frac{2(p_{2\mu}p_{1\nu}+p_{2\nu}p_{1\mu})}{S}
 \approx \frac{2(p_{2\mu}p_{1\nu}+p_{2\nu}p_{1\mu})}{S}
 \nonumber
\end{equation}
and the Sudakov parametrization of the  four-momenta of particles \cite{Arbuzov:2010zza},
\cite{Baier:1980kx}
\begin{equation}
q_i=\alpha_ip_2+\beta_ip_1+q_{i,t}, \ \  q_{i,t}p_{1,2}=0, \ \  q_{i,t}^2=-\vec{q_i^2}<0,
\nonumber
\end{equation}
we finally obtain
   \begin{equation}
   A_N=-\frac{5 \alpha_s\mu_a q_t(q_t^2+m_g^2)}{12 \pi M_q} \frac{ F_g(\rho
   |q_t|)N(q_t)}{D(q_t)},
   \label{SSA}
   \end{equation}
   where

  \begin{equation}
N(q_t)=\int \!\!d^2k_t \frac{(1+\mu_a^2(q_t \cdotp \! k_t + k_t^2)
   F_g(\rho |k_t|) F_g(\rho |q_t\!+\!k_t|)/(2M_q^2) }
  {(k_t^2+m_g^2)((k_t+q_t)^2+m_g^2)}\big)
\nonumber
\end{equation}
and
\begin{equation}
  D(q_t)= \Big(1+ (\frac{\mu_a q_t}{2M_q} F_g(\rho |q_t|))^2\Big)^2
    + \frac{\alpha_s^2 (q_t^2+m_g^2)^2}{12 \pi^2}
    \left( \int \!\! \frac{d^2k_t}{(k_t^2+m_g^2)((k_t+q_t)^2+m_g^2)}\right)^2 \nonumber
\end{equation}

\section{Results and discussion}
\begin{figure}[h]
\begin{minipage}[c]{8cm}
\vskip -0.5cm \hspace*{-1.0cm}
\centerline{\epsfig{file=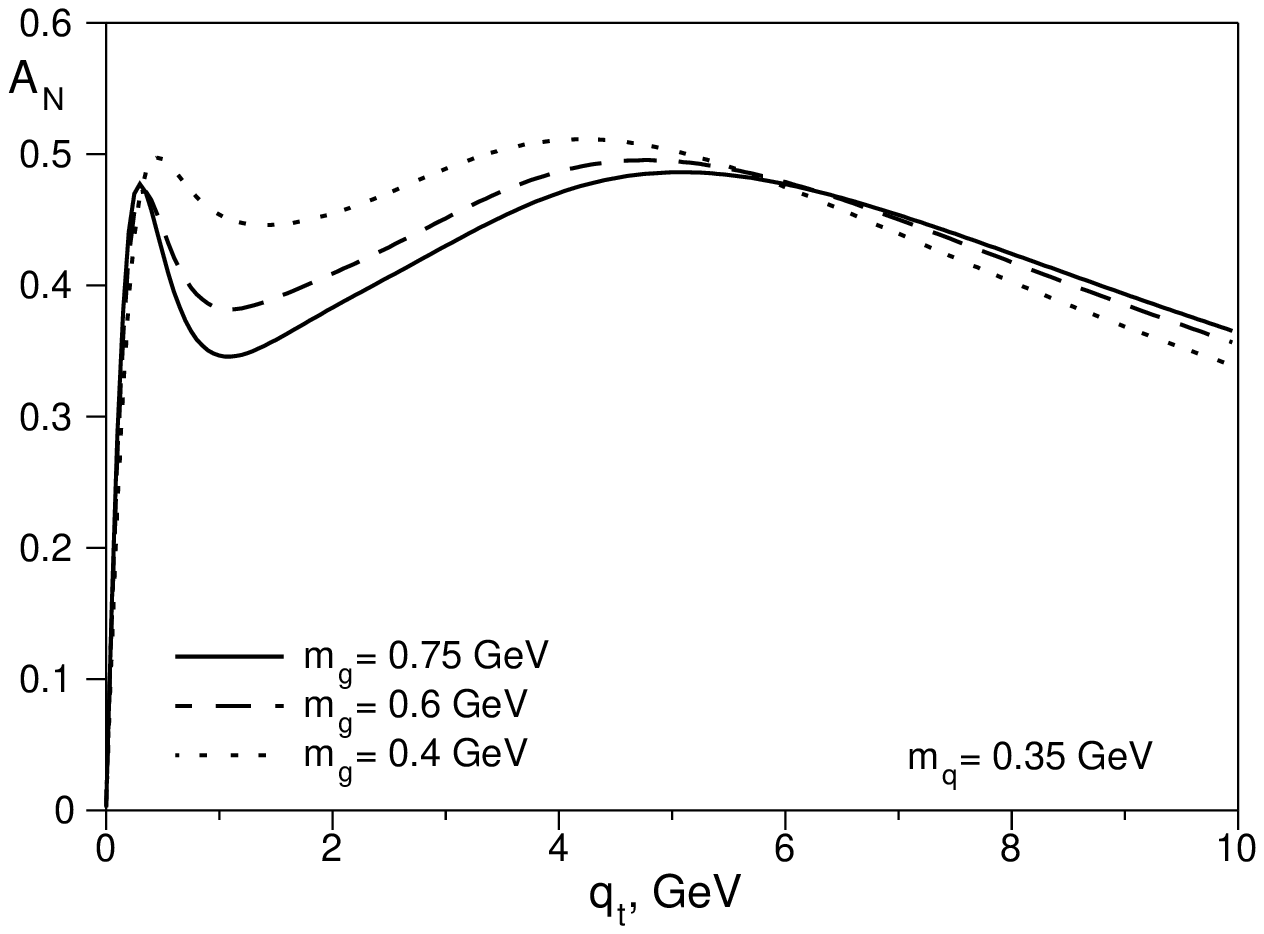,width=10cm,height=6cm,angle=0}}\
\end{minipage}
\begin{minipage}[c]{8cm}
\centerline{\epsfig{file=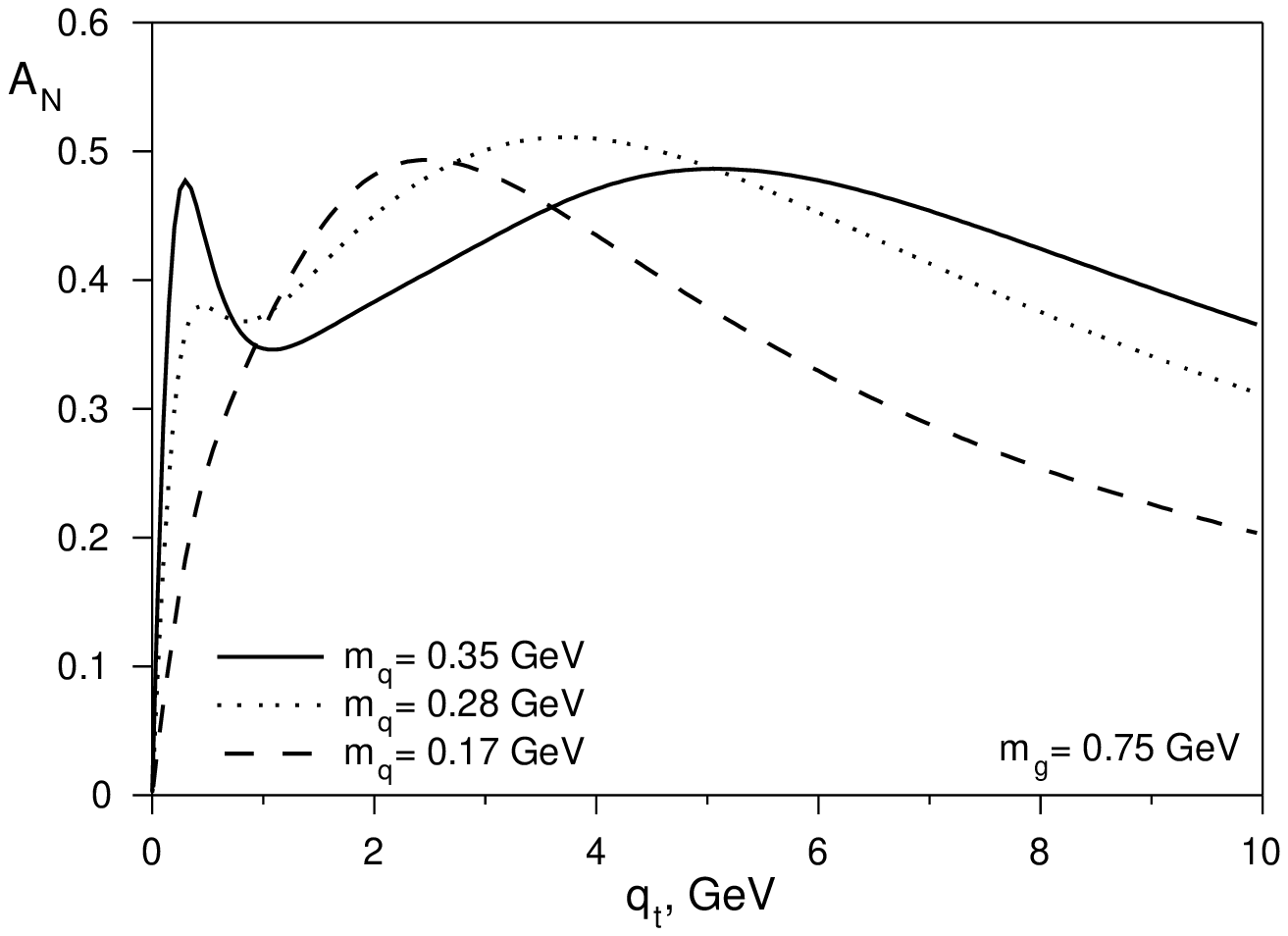,width=10cm,height=6cm,angle=0}}\
\hspace*{1.0cm} \vskip -1cm
\end{minipage}
\caption{ Left panel: the $q_t$ dependence of SSA for different
values of the infrared cut-off in the gluon propagator
\cite{nikolaev}, \cite{RuizArriola:2004en}, \cite{aguilar}. Right
panel: the $q_t$ dependence of  SSA for the different values of
the dynamical quark mass \cite{shuryak}, \cite{diakonov},
\cite{kochelev2}.}
\end{figure}
      In Fig.3, the result for $A_N$ as the function of transfer momentum transfer is presented
      for  different values of the dynamical quark mass $M_q$  and the parameter infrared cut-off
      $m_g$.
Our results show that SSA  $A_N$ induced by AQCM is very large and
practically independent of particular values of $M_q$ and $m_g$.
We would like to stress also that $A_N$ in our approach does not
depend on c.m. energy. The energy independence of  SSA is in
agreement with experimental data and in contradiction with naive
expectation that spin effects in strong interaction should vanish
at high energy \cite{Krisch:2010hr}.
 One can show that this property is directly related
to the spin one t-channel gluon exchange. Another remarkable
feature of our approach is a  flat dependence of SSA
 on transverse momentum of a final particle, Fig.3.
It comes from a rather soft power-like form factor in the
gluon-quark vertex, Eq.\ref{ffg},  and a small average size of
instanton, $\rho_c\approx 1/3 fm$, in QCD vacuum \cite{shuryak}.
Such a flat dependence has recently been observed by the STAR
collaboration in the inclusive $\pi^0$ production in high energy
proton-proton collision \cite{STAR} and was not expected in the
models based on TMD factorization and {\it ad hoc} parametrization
of Sivers and Collins functions \cite{reviews}. Finally, the sign
of the SSA is defined by the sign of AQCM and should be positive,
Eq.\ref{SSA}. This sign is very important in explaining of  the
signs of  SSA observed for inclusive production of $\pi^+,\pi^-$
and $\pi^0 $ mesons in proton-proton and proton-antiproton  high
energy collisions (see discussion and references in
\cite{reviews}, \cite{Krisch:2010hr}).

It is evident that the instanton induced helicity-flip should also
give the contribution   to SSA in the  meson production in
semi-inclusive deep inelastic scattering (SIDIS) where  large SSA
in $\pi$- and $K$-meson production was observed by HERMES
\cite{Airapetian:2010ds} and by COMPASS Collaborations
\cite{Martin:2013eja}. In the leading order in the instanton
density the nonzero contribution to SSA in SIDIS is expected to
come from the interference of diagrams presented in Fig.4. Here,
the imaginary part arises from final state perturbative and
nonperturbative interactions of the current quark with the
spectator system. The real part of the amplitude presented by two
first diagrams includes perturbative helicity-conserved
photon-quark vertex and the instanton induced helicity-flip
vertex. The Pauli form factor corresponding to the last vertex was
calculated in \cite{Kochelev:2003cp}.

\begin{figure}[htb]
\centering
   \centerline{\epsfig{file=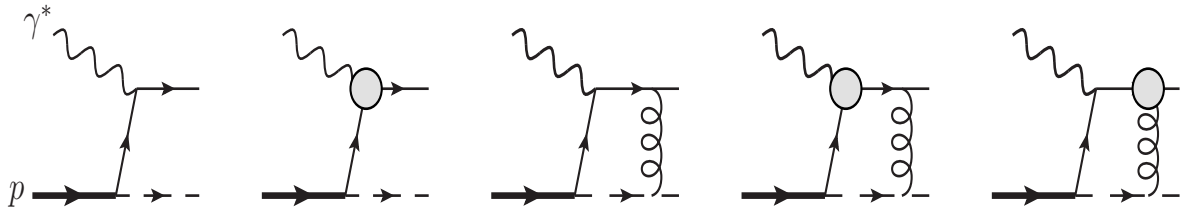,width=16cm,height=3cm,angle=0}}
    \caption{The leading contributions to SSA in SIDIS.}

   \end{figure}

 We should emphasize
the significant difference between our approach to SSA in SIDIS
and perturbative final state interaction model presented in
\cite{Brodsky:2002cx}. In particular,  one can expect that the
main contribution comes from the kinematical region where the
virtuality of gluon in Fig.4 is small. Therefore, soft gluon
interaction with quarks should be highly nonperturbative.
Furthermore, the helicity flip in \cite{Brodsky:2002cx} is related
to the wave function of the nucleon. Due to that, SSA coming from
this mechanism, might be significant only in the region of small
transverse momentum of the final meson $k_t\approx
\Lambda_{QCD}\approx 250$ MeV. In our approach, we expect the
large SSA at  higher transverse momentum because the averaged
instanton size is much smaller than the confinement size
$\rho_c\approx R_{conf}/3$. This qualitative observation
corresponds to the experimental data presented by HERMES and
COMPASS where large SSA was observed only at rather large $k_t$.
Additionally, a significant $Q^2$ dependence of SSA found by
COMPASS Collaboration  \cite{Martin:2013eja} might be related to
the strong $Q^2$ dependence of the nonperturbative photon-quark
vertex presented by second diagram in Fig.4.

 The additional contribution to SSA induced by instantons was
 suggested in the papers \cite{Ostrovsky:2004pd} and \cite{Qian:2011ya}.
It is based  on the results from \cite{Moch:1996bs},
 where the effects of instantons in
the nonpolarized deep inelastic scattering process  were
calculated in a careful way \footnote{This approach was applied to
the Drell-Yan process \cite{Brandenburg:2006xu} as well.}.  In
this case, the effect arises from phase shift in the quark
propagator in the instanton field. This contribution might be
considered as complementary to the AQCM effect.

In spite of the fact that our estimation is based mainly on
single-instanton approximation (SIA) for AQCM \cite{kochelev1},
the effects of the multiinstantons, which are hidden in the value
of dynamical quark mass in  Eq.\ref{AQCM}, are also taken into
account in the effective way. The accuracy of such SIA was
discussed in various aspects in \cite{Faccioli:2001ug}. By
analyzing of several correlation functions the authors claimed
that dynamical quark mass can be different from the MFA value
$M_q=170$ MeV. However, as it was discussed above, SSA induced by
AQCM has rather a weak dependence on the value of dynamical mass,
Fig.3. Therefore, we believe that some effects beyond SIA can not
lead to a significant change of our results.
 Furthermore, we would like to mention that the SSA
mechanism based on AQCM is quite general and might  happen in any
nonperturbative QCD model with the spontaneous chiral symmetry
breaking. The attractive feature of the  instanton liquid model is
that within this model  this phenomenon comes from rather small
distances $\rho_c\approx 0.3$ fm.  As the result, it  allows  to
understand the origin  of   large observed  SSA at large
transverse momentum.

In summary, we calculated the SSA in the quark-quark scattering
induced by AQCM and found that it was large. This phenomenon  is
related  to the strong  helicity-flip  quark-gluon interaction
induced by the topologically nontrivial configuration of vacuum
gluon fields called instantons. Our estimation shows that the
suggested mechanism can be responsible for anomalously large SSA
observed in different reactions at high energies. We would like to
stress that quark-gluon and quark-photon nonperturbative
interactions violate the TMD factorization in inclusive meson
production in both hadron-hadron and  deep inelastic scatterings.
Therefore, it cannot be treated as some additional contribution to
the Sivers distribution function or to the Collins fragmentation
function. It is evident that the nonfactorizable mechanism for SSA
based on AQCM can be extended to other spin-dependent observables,
including double-spin asymmetries in inclusive and exclusive
reactions.

\section*{Acknowledgments}
The authors are very grateful to  I.O. Cherednikov, A.E. Dorokhov,
A.V. Efremov and E.~A.~Kuraev for discussion. The work of N.K. was
supported in part by Belarus-JINR grant, a visiting scientist
grant from the University of Valencia and  by the MEST of the
Korean Government (Brain Pool Program No. 121S-1-3-0318). We also
acknowledge that this work was initiated through the series of
APCTP-BLTP JINR Joint Workshops.

\end{document}